\documentclass[12pt]{article}
\usepackage{amsmath}
\usepackage{graphicx}
\usepackage{amsfonts}
\usepackage{amssymb}
\newenvironment{proof}[1][Proof]{\textbf{#1.} }{\ \rule{0.5em}{0.5em}}

\begin{document}

\title{Laurent Expansions for Vertex Operators }
\date{}
\author{Wojtek Slowikowski\\Department of Mathematical Sciences, Aarhus University, Denmark}
\maketitle
\begin{abstract}
A method is presented for using coherent vectors to calculate the explicit
form of Schur polynomials which are the coefficients of Laurent expansion of a
vertex operator.
\end{abstract}

\section{Preliminaries}

Let $\Gamma_{0}\mathcal{D}$ be a Bose algebra (cf. \cite{Nielsen}) i.e. a
commutative graded algebra generated by a pre-Hilbert space $\mathcal{D}%
,\left\langle ,\right\rangle $ (the so-called one-particle space) and the
unity $\phi$ (the vacuum) provided with the extension $\left\langle
,\right\rangle $ of the scalar product of $\mathcal{D}$ making $\phi$ a unit
vector and fulfilling the property that for every $x\in\mathcal{D}$, the
adjoint $x^{\ast}$ to the operator of multiplication by $x$ is defined on the
whole $\Gamma_{0}\mathcal{D}$ and constitutes a derivation (i.e. fulfils the
Leibniz rule). We make the space $\widetilde{\Gamma}\mathcal{D}$ of all
antilinear functionals on $\Gamma_{0}\mathcal{D}$ the extension of $\Gamma
_{0}\mathcal{D}$ by identifying $f\in\Gamma_{0}\mathcal{D}$ with the
antilinear functional $\left\langle \cdot,f\right\rangle .$ The space
$\widetilde{\Gamma}\mathcal{D}$ can be naturally made into an algebra
containing $\Gamma_{0}\mathcal{D}$ as a subalgebra. We consider $\widetilde
{\Gamma}\mathcal{D}$ as a locally convex space with the weak topology
$\sigma\left(  \widetilde{\Gamma}\mathcal{D},\Gamma_{0}\mathcal{D}\right)  .$
The weak closure $\widetilde{\mathcal{D}}$ of $\mathcal{D}$ is a subspace of
$\widetilde{\Gamma}\mathcal{D}.$ It is easy to show that $\Gamma
_{0}\mathcal{D},\left\langle ,\right\rangle $ admits the completion
$\Gamma\overline{\mathcal{D}}$ within $\widetilde{\Gamma}\mathcal{D}$.

We shall use the exponentials of elements $w\in\mathcal{D},$%
\[
e^{w}=\sum_{n=0}^{\infty}\frac{1}{n!}w^{n}\in\Gamma\overline{\mathcal{D}},
\]
which are called \emph{coherent vectors}. In \cite{Nielsen} the following
relations are verified:
\begin{equation}
\left\langle a,b\right\rangle ^{j}=\frac{1}{j!}\left\langle a^{j}%
,b^{j}\right\rangle \label{sc-pro-of-powers}%
\end{equation}%
\begin{equation}
\left(  x^{n}\right)  ^{\ast}e^{w}=\left\langle x,w\right\rangle ^{n}e^{w}
\label{deriv-of-exp}%
\end{equation}%
\begin{equation}
\left\langle e^{u},fg\right\rangle =\left\langle e^{u},f\right\rangle
\left\langle e^{u},g\right\rangle \label{multiplicability}%
\end{equation}%
\begin{equation}
e^{\mathbf{a}\left(  w\right)  }e^{v}=e^{\left\langle w,v\right\rangle }e^{v}.
\label{exp-der-of-exp}%
\end{equation}
Also a proof that the set $\left\{  e^{x}:x\in\mathcal{D}\right\}  $ of
coherent vectors is total in $\Gamma\overline{\mathcal{D}}$ can be found in
\cite{Nielsen}.

\section{The Laurent Expansion for a Vertex operator}

Let $\mathcal{D}$ be spanned by an orthonormal system $\left\{  f_{n}\right\}
$ and by an orthonormal system $\left\{  g_{n}\right\}  $ as well. The
operator valued functions of $z$
\[
V\left(  z\right)  =e^{\sum_{n=1}^{\infty}z^{n}f_{n}}e^{\sum_{n=1}^{\infty
}z^{-n}g_{n}^{\ast}}:\Gamma_{0}\mathcal{D}\rightarrow\widetilde{\Gamma
}\mathcal{D,}%
\]
shall be called a \emph{vertex operator }(cf.\cite{Jing},\cite{Kac}%
,\cite{monster}).

Write $\left(  \frak{p,q}\right)  $ for tuples of non-negative integers
\[
\left(  \frak{p,q}\right)  =\left(  p_{1},q_{1},p_{2},q_{2},\cdots,p_{k}%
,q_{k},\cdots\right)
\]
and define
\[
\frak{N}_{m}=\left\{  \left(  \frak{p,q}\right)  :\sum_{k=1}^{\infty}\left(
p_{k}+q_{k}\right)  =m\right\}
\]
and
\[
\frak{N}^{w}=\left\{  \left(  \frak{p,q}\right)  :\sum_{k=1}^{\infty}\left(
p_{k}+q_{k}\right)  <\infty,\;\sum_{j=1}^{\infty}j\left(  p_{j}-q_{j}\right)
=w\right\}  .
\]
For $\frak{s=}\left(  s_{1},s_{2},...\right)  ,$ write
\[
\frak{s!=}\prod_{k=1}^{\infty}s_{k}!.
\]
We prove the following

\smallskip\ 

THEOREM \textit{Vertex operators admit the weak evaluation on} $\Gamma
_{0}\mathcal{D}$ \textit{and the weak convergent Laurent expansion}
\[
V\left(  z\right)  =e^{\sum_{n=1}^{\infty}z^{n}f_{n}}e^{\sum_{n=1}^{\infty
}z^{-n}g_{n}^{\ast}}=\sum_{w\in\mathbb{Z}}\mathcal{S}_{w}\left\{  f_{n}%
,g_{n}^{\ast}\right\}  z^{w}%
\]
\textit{with coefficients}
\[
\mathcal{S}_{w}\left\{  f_{n},g_{n}^{\ast}\right\}  =\sum_{m=0}^{\infty}%
\sum_{\left(  \frak{p,q}\right)  \in\frak{N}_{m}\cap\frak{N}^{w}}\frac
{1}{\frak{p}!\frak{q!}}\left(  \prod_{k=1}^{\infty}f_{k}^{p_{k}}\right)
\left(  \prod_{k=1}^{\infty}g_{k}^{q_{k}}\right)  ^{\ast}%
\]
\textit{called the Schur polynomials} (cf.\cite{Kac}).

To prove the Theorem we shall need the following

LEMMA \textit{Take any pair of elements} $u,v\in\mathcal{D}.$ \textit{Then the
element} $V\left(  z\right)  e^{u}$ \textit{is well defined in} $\widetilde
{\Gamma}\mathcal{D}$ \textit{and we have}
\begin{equation}
\left\langle e^{u},V\left(  z\right)  e^{v}\right\rangle =\left\langle
e^{u},\left(  \sum_{w\in\mathbb{Z}}\mathcal{S}_{w}\left\{  f_{n},g_{n}^{\ast
}\right\}  z^{w}\right)  e^{v}\right\rangle , \label{exp,exp}%
\end{equation}
\textit{where}
\[
\mathcal{S}_{w}\left\{  f_{n},g_{n}^{\ast}\right\}  =\sum_{m=0}^{\infty}%
\sum_{\left(  \frak{p,q}\right)  \in\frak{N}_{m}\cap\frak{N}^{w}}\frac
{1}{\frak{p}!\frak{q!}}\left(  \prod_{k=1}^{\infty}f_{k}^{p_{k}}\right)
\left(  \prod_{k=1}^{\infty}g_{k}^{q_{k}}\right)  ^{\ast}.
\]%

%TCIMACRO{\TeXButton{Proof}{\proof}}%
%BeginExpansion
\proof
%EndExpansion
. Take $u$,$v\in\mathcal{D}$. By virtue of (\ref{exp-der-of-exp}) we get
\[
\left\langle e^{u},e^{\mathbf{a}^{+}\left(  x\right)  }e^{\mathbf{a}\left(
y\right)  }e^{v}\right\rangle =e^{\left\langle u,v\right\rangle }%
e^{\left\langle u,x\right\rangle +\left\langle y,v\right\rangle },
\]
and consequently
\[
\left\langle e^{u},V\left(  z\right)  e^{v}\right\rangle =e^{\left\langle
u,v\right\rangle }e^{\sum_{n=1}^{\infty}\left(  \left\langle f_{n,}%
,u\right\rangle z^{n}+\left\langle v,g_{n}\right\rangle z^{-n}\right)  }.
\]
Since $u$ and $v$ are linear combinations of $f_{k}$ and $g_{k}\ $%
respectively, $\left\langle f_{n,},u\right\rangle z^{n}=\left\langle
v,g_{n}\right\rangle z^{-n}=0$ for large $n$. Due to (\ref{deriv-of-exp}) we
get
\begin{align*}
&  \left\langle e^{u},\left(  \prod_{k=1}^{\infty}f_{k}^{p_{k}}\right)
\left(  \prod_{k=1}^{\infty}g_{k}^{q_{k}}\right)  ^{\ast}e^{v}\right\rangle \\
&  =\left\langle \left(  \prod_{k=1}^{\infty}f_{k}^{p_{k}}\right)  ^{\ast
}e^{u},\left(  \prod_{k=1}^{\infty}g_{k}^{q_{k}}\right)  ^{\ast}%
e^{v}\right\rangle =\left(  \prod_{k=1}^{\infty}\left\langle f_{k}%
,u\right\rangle ^{p_{k}}\left\langle v,g_{k}\right\rangle ^{q_{k}}\right)
e^{\left\langle u,v\right\rangle },
\end{align*}
where all the products are finite and they are non-zero only when $p_{k}$ and
$q_{k}$ are zeros for $f_{k}$ and $g_{k}$ orthogonal to $v$ and $u$
respectively. Consequently
\begin{align*}
&  \frac{1}{m!}\left(  \sum_{n=1}^{\infty}\left\langle f_{n,},u\right\rangle
z^{n}+\sum_{n=1}^{\infty}\left\langle v,g_{n}\right\rangle z^{-n}\right)
^{m}\\
&  =\sum_{\left(  \frak{p,q}\right)  \in\frak{N}_{m}}\frac{1}{\frak{p}%
!\frak{q!}}\prod_{k=1}^{\infty}\left(  \left\langle f_{k},u\right\rangle
^{p_{k}}\left\langle v,g_{k}\right\rangle ^{q_{k}}z^{k\left(  p_{k}%
-q_{k}\right)  }\right) \\
&  =\sum_{w\in\mathbb{Z}}\sum_{\left(  \frak{p,q}\right)  \in\frak{N}^{w}%
\cap\frak{N}_{m}}\frac{1}{\frak{p}!\frak{q!}}\left(  \prod_{k=1}^{\infty
}\left\langle f_{k},u\right\rangle ^{p_{k}}\left\langle v,g_{k}\right\rangle
^{q_{k}}\right)  z^{w}\\
&  \left\langle e^{u},\sum_{w\in\mathbb{Z}}\left(  \sum_{\left(
\frak{p,q}\right)  \in\frak{N}^{w}\cap\frak{N}_{m}}\frac{1}{\frak{p}%
!\frak{q!}}\left(  \prod_{k=1}^{\infty}f_{k}^{p_{k}}\right)  \left(
\prod_{k=1}^{\infty}g_{k}^{q_{k}}\right)  ^{\ast}\right)  z^{w}e^{v}%
\right\rangle e^{-\left\langle u,v\right\rangle }.
\end{align*}
Hence
\begin{align*}
&  \frac{1}{m!}\left(  \sum_{n=1}^{\infty}\left\langle f_{n,},u\right\rangle
z^{n}+\sum_{n=1}^{\infty}\left\langle v,g_{n}\right\rangle z^{-n}\right)
^{m}\\
&  =\left\langle e^{u},\sum_{w\in\mathbb{Z}}\left(  \sum_{\left(
\frak{p,q}\right)  \in\frak{N}_{m}\cap\frak{N}^{w}}\frac{1}{\frak{p}%
!\frak{q!}}\left(  \prod_{k=1}^{\infty}f_{k}^{p_{k}}\right)  \left(
\prod_{k=1}^{\infty}g_{k}^{q_{k}}\right)  ^{\ast}\right)  z^{w}e^{v}%
\right\rangle e^{-\left\langle u,v\right\rangle },
\end{align*}
and finally
\begin{align*}
\left\langle e^{u},V\left(  z\right)  e^{v}\right\rangle  &  =e^{\left\langle
u,v\right\rangle }e^{\sum_{n=1}^{\infty}\left(  \left\langle f_{n,}%
,u\right\rangle z^{n}+\left\langle v,g_{n}\right\rangle z^{-n}\right)  }\\
&  =\left\langle e^{u},\sum_{m=0}^{\infty}\sum_{w\in\mathbb{Z}}\left(
\sum_{\left(  \frak{p,q}\right)  \in\frak{N}_{m,w}}\frac{1}{\frak{p}%
!\frak{q!}}\left(  \prod_{k=1}^{\infty}f_{k}^{p_{k}}\right)  \left(
\prod_{k=1}^{\infty}g_{k}^{q_{k}}\right)  ^{\ast}\right)  z^{w}e^{v}%
\right\rangle
\end{align*}
which concludes the proof of the Lemma.

\smallskip\ 

\textbf{Proof of the Theorem}

Since $\Gamma_{0}\mathcal{D}$ is the linear span of the set $\left\{
x^{k}:x\in\mathcal{D},\;k=1,2,..\right\}  $ (cf.\cite{Nielsen})$,$ it is
sufficient to show that for any $u,v\in\mathcal{D}$ and any natural numbers
$k,j$ we have
\[
\left\langle u^{k},V\left(  z\right)  v^{j}\right\rangle =\left\langle
u^{k},\left(  \sum_{w\in\mathbb{Z}}\mathcal{S}_{w}\left\{  f_{n},g_{n}^{\ast
}\right\}  z^{w}\right)  v^{j}\right\rangle
\]
which follows by differentiating respectively $k$ and $j$ times at $0$ the
variables $t$ and $s$ of the identity (\ref{exp,exp}) with $tu$ and $sv$
substituted for $u$ and $v.$%
%TCIMACRO{\TeXButton{End Proof}{\endproof}}%
%BeginExpansion
\endproof
%EndExpansion

\end{document}